\documentclass[conference,10pt]{IEEEtran}

\IEEEoverridecommandlockouts

\usepackage{cite}
\usepackage{amsmath,amssymb}
\usepackage{graphicx}
\usepackage{booktabs}
\usepackage{multirow}
\usepackage{algorithm}
\usepackage{algorithmic}
\usepackage{array}
\usepackage{threeparttable}
\usepackage{pifont}
\usepackage{url}
\usepackage[hidelinks]{hyperref}

\newcommand{\cmark}{\ding{51}}
\newcommand{\xmark}{\ding{55}}

\graphicspath{{figures/}}
\renewcommand{\arraystretch}{0.92}

\begin{document}

\title{CEMA-Router: Infrastructure-Aware Batch Routing for LLM Inference in Cloud--Edge Systems}

\author{
  \IEEEauthorblockN{
    Xijia Wang,
    TianZhang He,
    Mohammad Goudarzi\textsuperscript{*}
  }
  \IEEEauthorblockA{
    Faculty of Information Technology, Monash University, Melbourne, Australia
  }
  \thanks{%
    \textsuperscript{*}Corresponding author\newline
    Emails: wangxijia99@gmail.com,
    tianzhang.he@monash.edu,
    mohammad.goudarzi@monash.edu.
  }
}

\maketitle

\begin{abstract}
Large language models (LLM) are increasingly integrated into pervasive applications spanning heterogeneous cloud and edge resources. 
These heterogeneous resources differ in computational capacity, energy efficiency, network conditions, and deployed model quality, creating trade-offs among response quality, monetary cost, energy consumption, and batch completion time when routing inference requests. Existing LLM routers typically make per-query routing decisions based primarily on model quality and token cost, whereas cloud–edge schedulers often overlook quality differences among deployed models. Consequently, these objectives remain insufficiently coordinated. We present CEMA-Router, an infrastructure-aware batch router for LLM inference in pervasive cloud–edge systems. CEMA-Router formulates batch routing as a multi-objective assignment problem that jointly minimizes model-quality loss, monetary cost, energy consumption, and straggler-dominated batch completion time. To efficiently explore the assignment space, it employs a routing-specific memetic algorithm that combines largest-request-first, load-aware seeding with probabilistic, evaluator-guided Lamarckian repair. Routing decisions are informed by deployment-specific node profiles, dynamic backlog estimates, and model-quality signals. We further implement CEMA-Router as an end-to-end system comprising a request manager, routing service, and profiler, supporting
both real cloud--edge deployment and trace-driven simulation. We evaluate the system using RouterBench traces and GSM8K and MMLU workloads on a heterogeneous six-node cloud--edge testbed. Compared with state-of-the-art baselines, CEMA-Router reduces the weighted infrastructure-aware objective by up to 86.3\% for GSM8K and 80.5\% for MMLU, while reducing energy consumption by up to 33.8\% and 19.5\% and batch
completion time by up to 27.1\% and 30.1\%, respectively. Quality-oriented baselines remain useful reference points for raw response quality, while CEMA-Router explicitly exposes the stronger deployment policy by jointly accounting for quality, cost, energy, and batch completion time.
\end{abstract}


\section{Introduction}
LLM-backed applications are rapidly transitioning from centralized interactive services toward pervasive deployments that span high-performance cloud GPUs and resource-constrained edge nodes close to users. While this shift offers benefits in wide-area latency reduction and data privacy, it introduces a highly complex routing problem. Existing LLM routers typically focus on model-level quality and token cost in isolation, abstracting away the real infrastructure. Consequently, they fail to account for critical deployment constraints such as hardware resource utilization, node-level queuing delay, network bandwidth, and energy consumption.

To address this gap, this paper introduces CEMA-Router, an infrastructure-aware LLM request routing system formulated as a batch-level assignment problem. Rather than making myopic, per-query decisions, CEMA-Router jointly routes short batches of requests. It optimizes a comprehensive multi-objective function encompassing profiled quality loss, explicit monetary cost, energy consumption, and batch completion time. To solve this NP-hard assignment efficiently online, we design a customized Memetic Algorithm (MA) equipped with continuous-batching queue models and a lightweight Retrieval-Augmented Generation (RAG) quality predictor.

\begin{figure}[t]
    \centering
    \includegraphics[width=0.95\columnwidth]{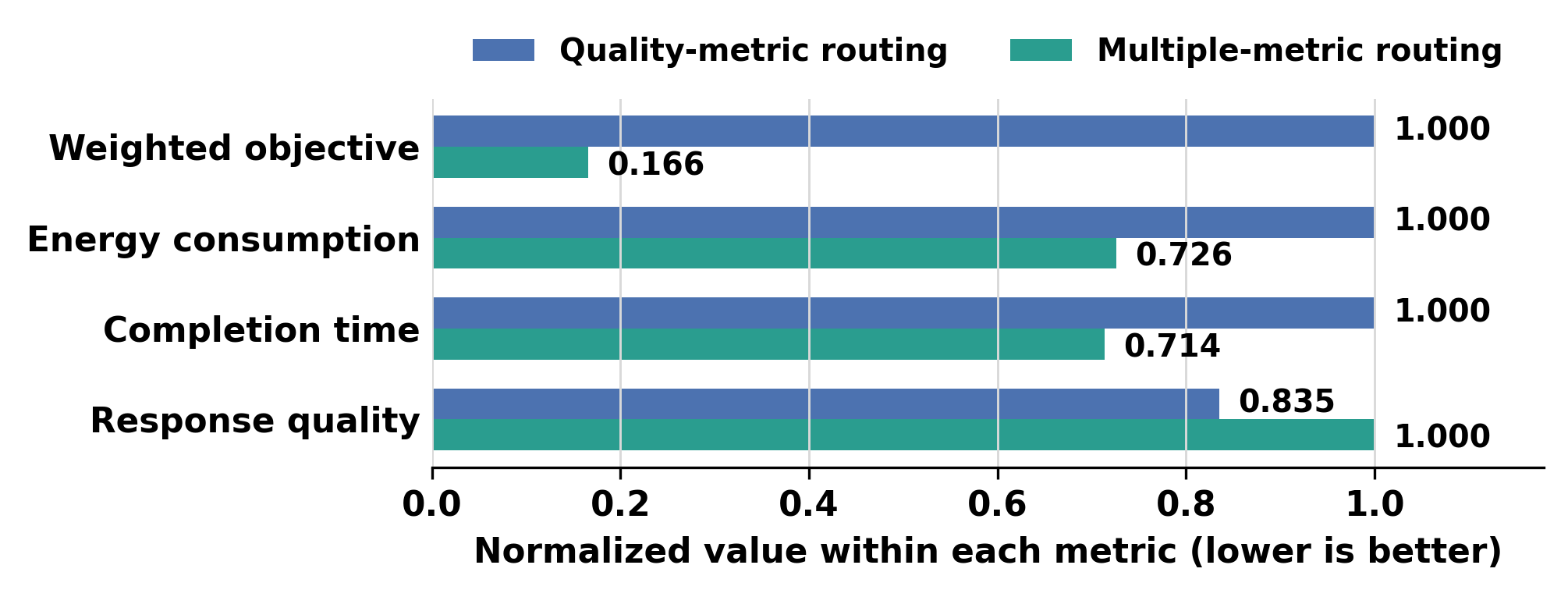}
    \caption{Quality-metric routing vs. multiple-metric routing in the existing six-node deployment. Values are normalized within each metric for readability.}
    \label{fig:intro-preview}
\end{figure}

The term infrastructure-aware captures the vital interaction between model capability and node capacity: the most accurate model becomes the wrong target if its hosting node is a bottleneck, or if another endpoint provides a superior latency-energy trade-off. A multiple-metric objective significantly improves the overall operational efficiency, whereas conventional quality-metric routing incurs latency and energy penalties to achieve only marginal raw quality gains (Fig.~\ref{fig:intro-preview}).

\begin{figure}[t]
    \centering
    \includegraphics[width=0.8\columnwidth]{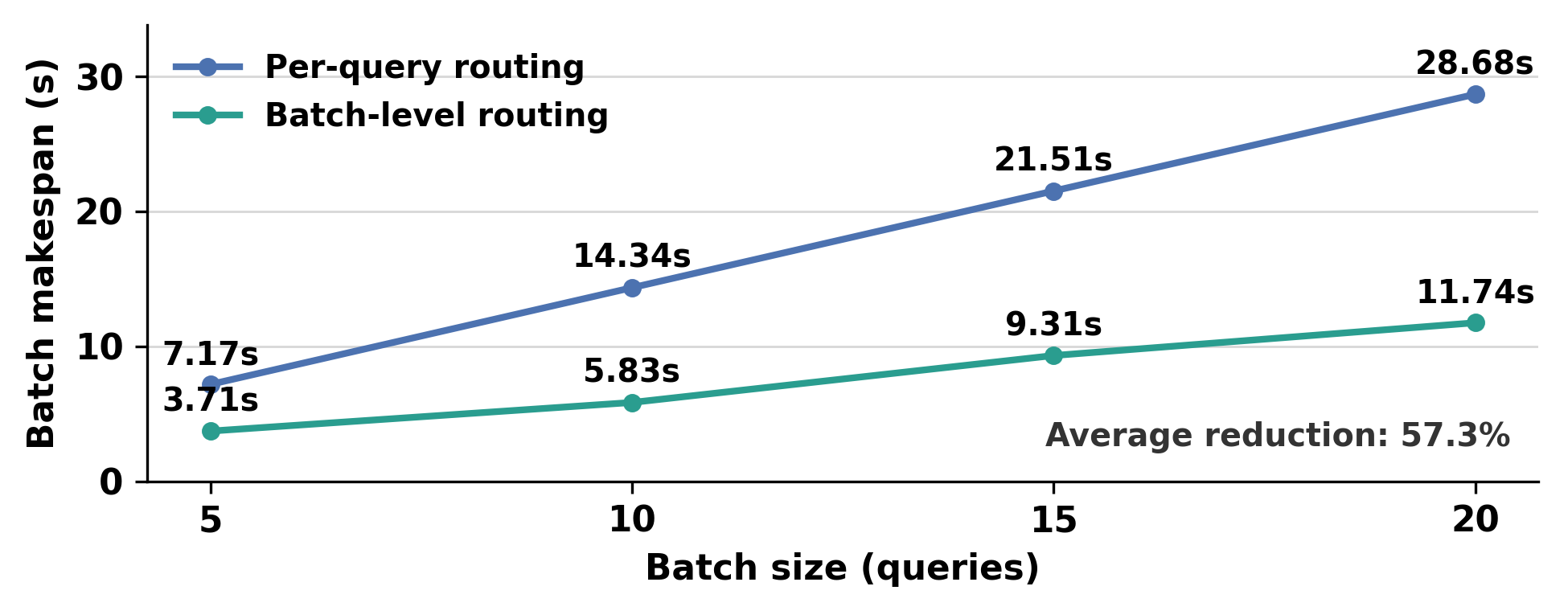}
    \caption{Per query vs batch LLM routing.}
    \label{fig:intro-batch}
\end{figure}

Also, to mitigate inference contention, Fig.~\ref{fig:intro-batch} reports a micro-benchmark validating our batch-level intuition. When requests are routed independently, multiple queries often crowd the same locally attractive endpoint, causing severe queuing delays. However, joint batch routing orchestrates globally and jointly assigns query batches and reduces average makespan by 57.3\%. 

We ask whether a deployable LLM router can leverage dynamic node states and RAG-based quality profiles to produce superior cloud-edge assignments while maintaining sub-second decision overhead. To answer this, we evaluate CEMA-Router through a synthesis of controlled RouterBench~\cite{hu2024routerbench} traces and a real six-node cloud-edge testbed running GSM8K~\cite{cobbe2021training} and MMLU~\cite{hendrycks2021measuring} workloads. This methodology isolates algorithmic performance while validating the end-to-end request management and profiling paths.

The main contributions are:
\begin{itemize}
    \item We formulate the LLM inference routing problem as a multi-objective optimization problem that jointly considers inference quality, monetary cost, energy consumption, and response latency, incorporating a FIFO queuing model to capture realistic request scheduling delays in cloud-edge environments.
    \item We design a customized Memetic Algorithm-based routing framework that modifies the MA pipeline for cloud-edge LLM serving through routing-aware chromosome encoding, mixed heuristic-random initialization, analytical multi-objective fitness evaluation, probabilistic Lamarckian crossover, and domain-specific local search operators including worst-gene repair and 2-opt swap.
    \item We develop a RAG-based quality prediction module that estimates request-model quality from similar historical prompts using sentence-transformer embeddings and FAISS indexing, avoiding trial inference or task-specific predictor training during routing, allowing quality to be combined with real-time infrastructure metrics.
    \item An evaluation using RouterBench traces and real GSM8K/MMLU workloads on a
    six-node cloud--edge deployment, showing that CEMA-Router reduces the weighted
    infrastructure-aware objective relative to UniRoute and simple deployment
    policies while exposing the quality--infrastructure trade-off.
\end{itemize}

\section{Background and Related Work}
\subsection{LLM Serving and Model Routing}

LLM serving focuses on optimizing the execution of trained language models under stringent memory, compute, and latency constraints. The Transformer architecture~\cite{vaswani2017attention} underlies modern LLMs such as GPT-3~\cite{brown2020gpt3}, where each request consumes accelerator memory, prefill/decode compute, and key-value (KV) cache capacity. State-of-the-art systems such as ORCA~\cite{yu2022orca}, Sarathi-Serve~\cite{agrawal2024taming}, DistServe~\cite{zhong2024distserve}, Splitwise~\cite{patel2024splitwise}, and vAttention~\cite{prabhu2025vattention} have made major progress in batching, phase disaggregation, and memory management. However, these systems excel at optimizing the execution path only after a request has already reached a specific model endpoint.

Conversely, model routing addresses the decision of which model or provider should process a request before execution begins. Recent frameworks such as RouteLLM~\cite{ong2024routellm}, Hybrid LLM~\cite{ding2024hybrid}, PolyRouter~\cite{stripelis2024polyrouter}, RouterBench~\cite{hu2024routerbench}, and UniRoute~\cite{jitkrittum2025universal} offer preference-based, quality-cost, or learned routing policies. Other routers employ tabular adaptation, ranking models, and mathematical optimization~\cite{jha2024learned,wang2023tabi,zhao2024eagle,feng2024graphrouter,mei2025omnirouter,vsakota2024fly,ding2025best,li2025llm,pan2025route,zhang2025query,song2025irt}. Crucially, this line of work primarily focuses on per-query model-level quality and token-cost trade-offs, often treating the underlying infrastructure as an abstracted, infinite-capacity black box. They typically ignore how routing a batch of requests to the same destination affects local queuing delays or real energy consumption.
\subsection{Cloud--Edge and Pervasive AI Systems}

Cloud--edge computing provides a distributed deployment model where cloud servers, nearby edge nodes, and user-side environments jointly host application services. Pervasive AI systems leverage this architecture to bring inference closer to data sources, thereby reducing wide-area latency and preserving privacy~\cite{shi2016edge,satyanarayanan2017edge}. Recent pervasive systems increasingly emphasize runtime adaptation across heterogeneous devices and energy constraints. For instance, ECORE~\cite{ecore2026} routes deep learning requests to energy-conscious edge device-model pairs; MERCI~\cite{merci2026} adapts multi-expert inference for dynamic vision perception; and ERUN~\cite{edgefaas2026} optimizes memory-efficient Edge FaaS runtimes. These systems demonstrate that hardware heterogeneity, energy bounds, and runtime load are central concerns in pervasive deployments.

Recent cloud--edge LLM systems have begun to explore adaptive serving policies across heterogeneous deployment scales~\cite{yu2025efficient,jain2024intelligent,yang2024perllm,shi2025inferencedynamics,wang2025mixllm,mei2025helix}. They bring LLM inference closer to pervasive environments and motivate the need for routers that can seamlessly integrate model-level quality metrics with deployment-level resource states.

Table~\ref{tab:table_1} compares representative LLM routing and pervasive AI systems across deployment scales, request granularity, modeled heterogeneity, and target performance metrics (cost, quality, latency, and energy).
\begin{table*}[!t]
\centering
\footnotesize
\caption{Comparison of LLM routing Techniques}
\label{tab:table_1}
\begin{threeparttable}
\resizebox{0.95\textwidth}{!}{%
\setlength{\tabcolsep}{3pt}
\renewcommand{\arraystretch}{1.2}
\begin{tabular}{lcc c ccc l cc ccccc}
\toprule
\multirow{2}{*}{\textbf{Work}} &
\multicolumn{2}{c}{\textbf{Environment Architecture}} &
\multirow{2}{*}{\parbox{1.5cm}{\centering\textbf{Incoming} \\ \textbf{Request}}} &
\multicolumn{3}{c}{\textbf{Aspect of Heterogeneity}} &
\multirow{2}{*}{\textbf{Technique}} &
\multicolumn{2}{c}{\textbf{Scalability}} &
\multicolumn{5}{c}{\textbf{Performance Metrics}} \\
\cmidrule(lr){2-3} \cmidrule(lr){5-7} \cmidrule(lr){9-10} \cmidrule(lr){11-15}
 & \parbox{1.1cm}{\centering\textbf{Cloud} \\ \textbf{Nodes}}
 & \parbox{1.1cm}{\centering\textbf{Edge} \\ \textbf{Nodes}}
 &  & \parbox{1.1cm}{\centering\textbf{LLM} \\ \textbf{Model}}
 & \parbox{1.1cm}{\centering\textbf{LLM} \\ \textbf{Size}}
 & \parbox{1.5cm}{\centering\textbf{Computation} \\ \textbf{Nodes}}
 &  & \parbox{1.45cm}{\centering\textbf{Destination}}
 & \parbox{1.2cm}{\centering\textbf{Request}}
 & \textbf{Type}
 & \textbf{Cost}
 & \textbf{Quality}
 & \textbf{Latency}
 & \textbf{Energy} \\
\midrule
\cite{yu2025efficient} & Multiple & Multiple & Single & \cmark & \cmark & \cmark & Heuristic & \cmark & \cmark & System & \cmark & \cmark & \cmark & \xmark \\
\cite{jha2024learned} & Single & None & Single & \cmark & \cmark & \cmark & DQN & \xmark & \cmark & Hybrid & \xmark & \xmark & \xmark & \xmark \\
\cite{wang2023tabi} & Single & None & Single & \cmark & \cmark & \xmark & Heuristic & \xmark & \xmark & Hybrid & \xmark & \cmark & \cmark & \xmark \\
\cite{jain2024intelligent} & Multiple & None & Single & \xmark & \xmark & \cmark & DDQN & \xmark & \xmark & System & \xmark & \xmark & \cmark & \xmark \\
\cite{stripelis2024polyrouter} & Single & None & Single & \cmark & \cmark & \xmark & BERT-based & \xmark & \xmark & Hybrid & \cmark & \cmark & \xmark & \xmark \\
\cite{ding2024hybrid} & Single & None & Single & \cmark & \cmark & \cmark & BERT-based & \xmark & \xmark & Hybrid & \cmark & \cmark & \xmark & \xmark \\
\cite{zhao2024eagle} & Single & None & Single & \cmark & \cmark & \xmark & Elo-based & \xmark & \xmark & Hybrid & \xmark & \cmark & \xmark & \xmark \\
\cite{feng2024graphrouter} & Single & None & Single & \cmark & \cmark & \xmark & GNN-based & \xmark & \xmark & Hybrid & \cmark & \cmark & \xmark & \xmark \\
\cite{ong2024routellm} & Single & None & Single & \cmark & \cmark & \xmark & Ensemble-based & \cmark & \xmark & System & \cmark & \cmark & \xmark & \xmark \\
\cite{yang2024perllm} & Multiple & Multiple & Single & \cmark & \cmark & \cmark & Bandit-based & \cmark & \cmark & System & \xmark & \xmark & \cmark & \cmark \\
\cite{shi2025inferencedynamics} & Multiple & Multiple & Single & \cmark & \xmark & \xmark & Heuristic & \cmark & \cmark & ML & \cmark & \cmark & \xmark & \xmark \\
\cite{wang2025mixllm} & Multiple & Multiple & Single & \cmark & \cmark & \xmark & Bandit-based & \cmark & \xmark & System & \cmark & \cmark & \cmark & \xmark \\
\cite{mei2025omnirouter} & Single & None & Single & \cmark & \cmark & \xmark & Mathematical & \cmark & \cmark & Hybrid & \xmark & \cmark & \cmark & \xmark \\
\cite{vsakota2024fly} & Single & None & Single & \cmark & \cmark & \xmark & BERT-based & \xmark & \xmark & System & \cmark & \cmark & \xmark & \xmark \\
\cite{mei2025helix} & Multiple & None & Single & \xmark & \cmark & \cmark & Mathematical & \cmark & \cmark & System & \xmark & \xmark & \cmark & \xmark \\
\cite{ding2025best} & Single & None & Single & \cmark & \cmark & \xmark & BERT-based & \xmark & \xmark & Hybrid & \cmark & \cmark & \xmark & \xmark \\
\cite{li2025llm} & Single & None & Single & \cmark & \cmark & \xmark & Bandit-based & \xmark & \xmark & Hybrid & \cmark & \cmark & \xmark & \xmark \\
\cite{pan2025route} & Single & None & Single & \cmark & \cmark & \xmark & MLP-based & \xmark & \xmark & Hybrid & \xmark & \cmark & \xmark & \xmark \\
\cite{zhang2025query} & Single & None & Single & \cmark & \cmark & \xmark & Learning-based & \xmark & \xmark & Hybrid & \xmark & \cmark & \cmark & \xmark \\
\cite{song2025irt} & Single & None & Single & \cmark & \cmark & \xmark & IRT-based & \xmark & \xmark & Hybrid & \cmark & \cmark & \xmark & \xmark \\
\cite{jitkrittum2025universal} & Multiple & None & Single & \cmark & \cmark & \xmark & Heuristic & \cmark & \xmark & ML & \cmark & \cmark & \xmark & \xmark \\
CEMA-Router & Multiple & Multiple & Batch & \cmark & \cmark & \cmark & MA-based & \cmark & \cmark & Hybrid & \cmark & \cmark & \cmark & \cmark \\
\bottomrule
\end{tabular}
}
\vspace{2pt}
\begin{tablenotes}
\footnotesize
\item \parbox{\textwidth}{DQN = Deep Q-Network; DDQN = Double DQN; JCT = Job Completion Time; BERT = Bidirectional Encoder Representations from Transformers; \newline MLP = Multi-Layer Perceptron; GNN = Graph Neural Network; ML = Machine Learning; IRT = Item Response Theory; QNC = Quality-Neutral Cost.}
\end{tablenotes}
\end{threeparttable}
\end{table*}
\subsection{Positioning of CEMA-Router}

The comparison in Table~\ref{tab:table_1} identifies a critical gap in the current literature: existing solutions typically address only one side of the coin. LLM routers optimize for response quality and API cost, but completely abstract away the physical endpoint---ignoring node energy consumption, network path limits, and queuing bottlenecks induced by concurrent batch requests. Conversely, cloud--edge scheduling systems meticulously adapt to physical resource constraints but rarely incorporate prompt-specific LLM quality or token economics into their routing objectives.

CEMA-Router bridges this gap directly. It treats each deployed node--model pair as a concrete, resource-constrained destination. By combining an RAG-style quality predictor to estimate prompt-specific quality loss with token cost tables, calibrated endpoint throughput, network measurements, energy profiles, and real-time backlog states, CEMA-Router performs batch-level optimization. Consequently, it minimizes a unified, weighted objective (i.e. fitness) covering quality, monetary cost, energy consumption, and batch completion time, ensuring that every assignment respects both the semantic capability of the model and the physical limitations of the pervasive infrastructure.

\section{System Architecture and Routing Model}\label{sec:system}
\begin{figure}[!t]
    \centering
    \includegraphics[width=0.9\columnwidth, height=6cm]{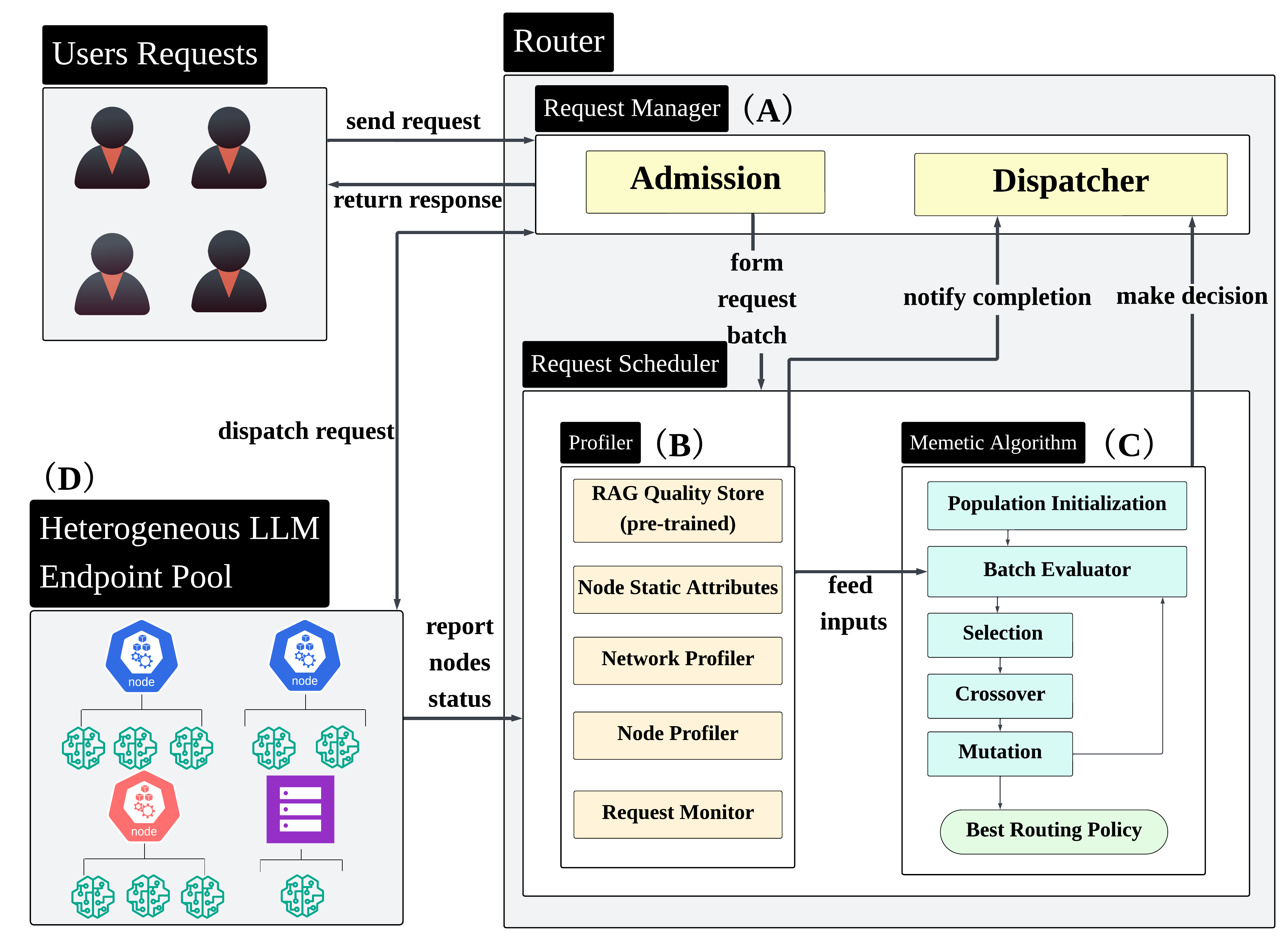}
    \caption{CEMA-Router architecture overview.}
    \label{fig:system-architecture}
\end{figure}

Figure~\ref{fig:system-architecture} presents CEMA-Router as a runtime routing
architecture with four main parts: the Request Manager, the Router, the Profiler,
and the heterogeneous LLM endpoint pool. The Request Manager admits user requests
and dispatches the selected routing policy. The Router contains the Request
Scheduler and the memetic algorithm that evaluates candidate batch assignments.
The Profiler groups the Request Monitor, Network Profiler, Node Profiler,
pre-trained RAG Quality Store, and Node Static Attributes, and feeds their routing
inputs to the BatchEvaluator. This section defines the system state and objective
used by the router; Section~\ref{sec:design} then explains how the memetic
algorithm searches over candidate assignments and obtains the best routing policy.

\subsection{Request Management}\label{subsec:request-management}
The Request Manager is the online entry and exit path for user workloads. Its
admission component operates in discrete routing windows $[T_0,T_1)$ of length
$\Delta$. All requests \(r\) arriving within the window are grouped into an admitted
batch $\mathcal{R}$, with size $R\le R_m$, and each request carries the fields
needed by the evaluator: prompt content, input-token count, predicted output-token
count, and network context. The admitted batch is forwarded to the Request
Scheduler. After the router returns the best routing policy, the dispatcher sends
each request to its assigned destination \(d\) in the heterogeneous endpoint pool.
Responses are returned through the same management path, while dispatch and
completion notifications are recorded by the Request Monitor inside the Profiler.

\subsection{Profiler and Routing Context}
\label{subsec:routing-context}

The Profiler provides the routing context consumed by the BatchEvaluator in
Section~\ref{subsec:scheduler-evaluator}. Static node attributes capture deployment
properties such as model placement, node type, token-price settings, calibrated
single-request and saturated-batch throughput, network goodput, and power
characteristics. The Node Profiler reports endpoint health and node status, while the
Network Profiler measures router-to-endpoint latency and network conditions. The
Request Monitor records dispatch and completion events and maintains the in-flight
backlog estimate $B_d$ for each destination as carry-over service time
from previously dispatched work. After each routing window, it updates $B_d$ by subtracting the elapsed window $\Delta$ from the previous backlog
and adding newly assigned service duration, retaining only unfinished work.
This system-side tracking is necessary because the endpoint vary in model family, model size, hardware architecture, and serving stack. CEMA-Router therefore maintans
a common in-flight view from its own dispatch and completion records rather than replying on endpoint-specific queue or telemetry interfaces.

The pre-trained RAG Quality Store is the quality predictor for routing. The router
must make an assignment quickly, so it cannot send the query to every candidate
endpoint and wait for observed response quality before choosing a destination. 
In the real deployment, the quality store is populated offline by profiling each workload across the available endpoints and scoring outputs automatically. Representative prompts are embedded with \texttt{all-MiniLM-L6-v2} and indexed in FAISS~\cite{johnson2019billion}. 
At runtime, the incoming prompt embedding retrieves the top-$K$ nearest
profiled prompts. For each destination $d$, the score is the
similarity-weighted average of the historical quality scores of those neighbors,
with higher-similarity prompts contributing more to $\hat{q}(r,d)$ (Eq.~\ref{eq:epred-sum}). The
BatchEvaluator then uses $1-\hat{q}(r,d)$ as the request--destination quality-loss
entry. The quality store provides prompt-specific quality estimates before dispatch without requiring trial inference.

The same retrieved quality evidence is supplied to CEMA-Router and UniRoute in the
deployment comparison. CEMA-Router combines this evidence with Request Monitor
backlog, calibrated batch throughput, network measurements, token cost, and energy
profiles, while UniRoute uses it mainly through its quality-cost objective.
Holding the quality source fixed isolates the effect of infrastructure-aware
assignment.

\subsection{Request Scheduler and Batch Evaluation}
\label{subsec:scheduler-evaluator}

The Request Scheduler passes each admitted batch and the latest Profiler context
to the Router. It assembles the request set, destinations, profiler signals, and objective weights, and then asks the memetic algorithm in Sect.~\ref{sec:design} to find the best destination per request. The BatchEvaluator scores complete batch assignments. For a batch 
$R$ and destinations $D$, an assignment maps each request to one concrete
node--model destination in the current endpoint pool. The evaluator computes following terms for each assignment. 

\textbf{Response-quality loss} $Q$ is computed by the similarity-weighted historical score from the RAG Quality: 

\begin{equation}
\label{eq:epred-sum}
  \hat{q}(r, d) =
  \frac{\sum_{i \in \mathcal{N}_K(r)} s_i \cdot q(p_i, d)}
       {\sum_{i \in \mathcal{N}_K(r)} s_i}.
\end{equation}

where $\mathcal{N}_K(r)$ is the retrieved neighbor set, $s_i$ is the cosine
similarity between $r$ and profiled prompt $p_i$, and $q(p_i,d)$ is the recorded
response-quality score of $p_i$ at destination $d$. If no neighbor exceeds the
similarity threshold, the predictor uses the destination's mean profiled quality.

\textbf{Monetary cost} $C$ combines destination
token-pricing entries with any metered transfer cost; in self-hosted deployment runs the external API price is zero for every endpoint.

\begin{equation}
C_{\mathrm{pred}} \;=\; C_{\mathrm{infer}} \;+\; C_{\mathrm{trans}}.
\label{eq:cpred-sum}
\end{equation}

\textbf{Energy consumption} $E$ is decomposed into processing and communication energy. Processing
energy uses the profiled service time of the selected node--model destination together with its board-power and utilization assumptions. Communication energy uses the upload and download payload sizes, link goodput, and transmit/receive power profile for the router-to-endpoint path. Thus the same model can contribute different energy under different node placements or network paths.

\begin{equation}
E_{\mathrm{pred}} \;=\; E_{\mathrm{infer}} \;+\; E_{\mathrm{trans}}.
\label{eq:epred-sum}
\end{equation}

\textbf{Completion time} $T$ is the batch makespan. For each request, the evaluator estimates inference time from model size, predicted token length, and calibrated single-request throughput; estimates transfer time from router-to-endpoint latency and uplink/downlink goodput; and estimates FIFO queuing from $B_d$ plus the cumulative work of earlier requests assigned to the same destination. 
\begin{equation}
\hat{T}_{r}
\;=\;
\hat{T}_{\mathrm{infer}}(r,d_r)
\;+\;
\hat{T}_{\mathrm{trans}}(r,d_r)
\;+\;
\hat{T}_{\mathrm{queue}}(r,d_r).
\label{eq:that-decompose}
\end{equation}
To account for continuous batching, the cumulative work is divided by the destination's saturated-batch speedup, while the prediction is bounded so a request is not treated as completing faster than its own service time. This makes the evaluator sensitive to interactions among requests: several individually attractive assignments can still create a poor batch if they concentrate work on the same endpoint and turn it into the straggler. For the backlog update after dispatch, the newly assigned service duration of a destination is approximated by the larger of its longest individual request service time and its total serial-equivalent work divided by the saturated-batch speedup. The next-window
backlog is then the unfinished remainder after subtracting $\Delta$. The saturated-batch throughput profile is used here rather than peak hardware FLOPS, so the service-time reflects the measured serving stack for the deployed model and precision.

\textbf{Weighted Cost} CEMA-Router’s primary objective is to minimize a weighted infrastructure-aware cost combining normalized response-quality loss, monetary cost, energy consumption, and batch completion time. Each weight reflects the relative importance of a metric to the deployment operator. By optimising these metrics jointly over the entire request batch, CEMA-Router selects the assignment that best satisfies the configured operational priorities rather than optimising any single metric in isolation.

\begin{equation}
J(x)=\omega_Q \widetilde{Q}(x)+\omega_C \widetilde{C}(x)+
\omega_E \widetilde{E}(x)+\omega_T \widetilde{T}(x),
\end{equation}
where $x$ is a complete request-to-destination assignment. Exact weights $\omega$ are part of the evaluation configuration. All terms are normalized to \([0,1]\) before aggregation,
\begin{equation}
    \mathrm{Metric}_{i,\mathrm{norm}}
    \;=\;
    \frac{\mathrm{Metric}_{i} - \mathrm{Metric}_{\min}}
         {\mathrm{Metric}_{\max} - \mathrm{Metric}_{\min}}.
\label{eq:normalize}
\end{equation}

\subsection{Heterogeneous Endpoint Pool}
\label{subsec:endpoint-pool}

The heterogeneous endpoint pool receives dispatcher assignments and executes LLM requests.  Operationally, if node $i$ hosts $M_i$ models, CEMA-Router flattens all hosted node--model pairs into a destination set $\mathcal{D}$. Destination-specific parameters, including token price, calibrated throughput, power, bandwidth, and latency, are keyed by that flattened destination. 

CEMA-Router uses a runtime-neutral endpoint abstraction based on dispatch, response, node-status, and completion records. Endpoints return responses to the Request Manager, report node status through the Node Profiler, and notify completion to the Request Monitor. The Profiler converts these observations into a
common routing context for the BatchEvaluator, giving the router consistent inputs
across GPU servers, edge devices, and remote endpoints with different runtime
software and telemetry support.

\begin{algorithm}[t]
    \caption{CEMA-Router Batch Assignment}
    \label{alg:cema-route}
    \scriptsize
    \begin{algorithmic}[1]
        \STATE \textbf{Input}: batch $\mathcal{R}$, destinations $\mathcal{D}$
        \STATE \textbf{Input}: context $\mathcal{C}$, weights $\omega$, search budget $(P,G)$
        \STATE \textbf{Output}: best assignment chromosome $g^*$ and decision metrics
        \STATE Remove unavailable destinations and build legal sets $\{\mathcal{D}_r\}$
        \STATE Load quality estimates, destination profiles, and backlogs $\{B_d\}$
        \STATE $\mathcal{P}\leftarrow\operatorname{InitPopulation}(\mathcal{R},\mathcal{D},\mathcal{C},P)$
        \FOR{$t=1$ to $G$}
            \STATE Evaluate $(Q,C,E,T)$ for each $g\in\mathcal{P}$ using BatchEvaluator
            \STATE Normalize metrics within $\mathcal{P}$ and compute scalar $J(g)$
            \STATE $\mathcal{P}'\leftarrow\emptyset$
            \WHILE{$|\mathcal{P}'|<P$}
                \STATE Select parents $g^{(1)},g^{(2)}$ by 3-tournament selection
                \STATE Create $h^{(1)},h^{(2)}$ with Algorithm~\ref{alg:cema-lamarckian}
                \STATE Mutate active genes of $h^{(1)},h^{(2)}$ with probability $p_m$
                \STATE Add $h^{(1)},h^{(2)}$ to $\mathcal{P}'$
            \ENDWHILE
            \STATE Evaluate $(Q,C,E,T)$ for each $h\in\mathcal{P}'$ using BatchEvaluator
            \STATE Normalize metrics within $\mathcal{P}\cup\mathcal{P}'$ and compute scalar $J$
            \STATE $\mathcal{P}\leftarrow$ best $P$ chromosomes from $\mathcal{P}\cup\mathcal{P}'$ by $J$
        \ENDFOR
        \STATE \textbf{Return} $g^*=\arg\min_{g\in\mathcal{P}}J(g)$
    \end{algorithmic}
\end{algorithm}

\section{CEMA-Router Design}
\label{sec:design}

This section describes how CEMA-Router turns the system context in
Fig.~\ref{fig:system-architecture} into a batch-level routing decision. Its goal is
to assign a whole request batch to heterogeneous cloud--edge destinations while
accounting for quality, token charge, energy, network delay, calibrated throughput,
and queue backlog. 
\subsection{Design Rationale}

CEMA-Router uses memetic search because the decision is both multi-objective and
assignment-dependent. With $R$ admitted requests and $M$ legal destinations, the
router faces $M^R$ possible assignments. The router must trade off profiled quality
loss, monetary cost, energy consumption, and completion time, while the
completion time term changes with the whole batch assignment: moving one request can
alter endpoint service duration, queueing delay, and makespan. CEMA-Router therefore
evaluates complete candidate assignments and captures interactions among requests in
the same batch.

Each candidate assignment is encoded as a chromosome. Each active gene corresponds to
one admitted request and stores its selected destination; inactive positions beyond
$R$ are ignored by the evaluator. Legal destinations come from the current deployment
state: unhealthy endpoints can be removed, destination profiles provide calibrated
throughput and power, the Request Monitor supplies $B_d$, and the quality store
supplies request--destination quality-loss estimates. The encoding is compact enough
for online routing while still supporting crossover, mutation, and local repair.

\subsection{Online Workflow}

Algorithm~\ref{alg:cema-route} gives the online workflow. For each batch, the router
materializes the context used by the BatchEvaluator: request features, destination
profiles, backlog estimates, profiled quality values, and objective weights. It then
initializes a population with Algorithm~\ref{alg:cema-init}, combining one load-aware
chromosome with random legal assignments for diversity. The load-aware chromosome
schedules longer requests first and assigns each to the currently earliest finishing
destination.

Each generation evaluates whole-batch assignments using BatchEvaluator, which computes and normalizes quality, cost, energy, and completion time
terms to obtain $J(g)$. 
Elitist selection retains the best assignments, and the router returns the best
assignment found within the search budget. Although tuning indicated that
population size and generation count can scale with batch size, the reported
experiments use the fixed configuration in Section~\ref{sec:evaluation}, making the
scalability results reflect workload size under a constant search budget.

Algorithm~\ref{alg:cema-init} makes this seeding step explicit. The first chromosome
uses request size, endpoint backlog, profiled inference time, and network delay before
the random chromosomes are added. The resulting initial population contains one
congestion-aware assignment for reducing stragglers together with random legal
assignments for exploration.

Ordering requests by predicted token length gives the largest requests first access
to all feasible destinations, before the fastest endpoint has accumulated a large
provisional queue. The earliest-finish score $A_d$ combines inherited backlog $B_d$,
profiled inference time $\tau_{r,d}$, and network delay $\nu_{r,d}$, so the seed
chromosome is guided by estimated completion impact under the current deployment
state.
$\hat{T}_{\mathrm{infer}}(r,d)$ and $\hat{T}_{\mathrm{trans}}(r,d)$ denote the evaluator's inference and network transmission time (Alg.~\ref{alg:cema-init}, Line 8), derived from throughput and network profiles.

\begin{algorithm}[t]
    \caption{Routing-Aware Population Initialization}
    \label{alg:cema-init}
    \scriptsize
    \begin{algorithmic}[1]
        \STATE \textbf{Input}: batch $\mathcal{R}$, legal destination sets $\{\mathcal{D}_r\}$
        \STATE \textbf{Input}: context $\mathcal{C}$, population size $P$
        \STATE \textbf{Output}: initial population $\mathcal{P}$
        \STATE $\mathcal{P}\leftarrow\emptyset$; initialize seed chromosome $g$
        \STATE Set provisional load $L_d\leftarrow B_d$ for each $d \in \mathcal{D}$
        \STATE Sort requests in $\mathcal{R}$ by predicted token length in descending order
        \FOR{each request $r$ in sorted order}
            \STATE For each $d\in\mathcal{D}_r$, estimate $A_d=L_d+\hat{T}_{\mathrm{infer}}(r,d)+\hat{T}_{\mathrm{trans}}(r,d)$
            \STATE Choose $d^*=\arg\min_{d\in\mathcal{D}_r}A_d$ and set $g_r\leftarrow d^*$
            \STATE Update $L_{d^*}\leftarrow L_{d^*}+\hat{T}_{\mathrm{infer}}(r,d^*)$
        \ENDFOR
        \STATE Add $g$ to $\mathcal{P}$
        \WHILE{$|\mathcal{P}|<P$}
            \STATE Add a random legal chromosome sampled from $\{\mathcal{D}_r\}$
        \ENDWHILE
        \STATE \textbf{Return} $\mathcal{P}$
    \end{algorithmic}
\end{algorithm}

\subsection{Memetic Search and Local Repair}

Algorithm~\ref{alg:cema-lamarckian} defines the probabilistic Lamarckian repair
operator used by CEMA-Router. After two-point crossover creates offspring, the router
may invoke a bounded repair step on the active request genes. The repair searches two
routing-specific neighborhoods. Worst-gene repair identifies the request whose
predicted completion time dominates the current makespan and tests alternative legal
destinations for that request. A 2-opt swap tests whether exchanging the destinations
of two requests reduces queue imbalance or energy cost. All trial moves are scored by
the current BatchEvaluator, and only the best improving move is written back into the
offspring. In memetic algorithm terminology, the operator is Lamarckian because the
locally improved genotype is returned to the population. Coupling repair to crossover
confines exploitation to newly recombined assignments and bounds the extra evaluator
calls added per offspring.

\begin{algorithm}[t]
    \caption{Probabilistic Lamarckian Crossover and Repair}
    \label{alg:cema-lamarckian}
    \scriptsize
    \begin{algorithmic}[1]
        \STATE \textbf{Input}: parents $g^{(1)},g^{(2)}$, legal sets $\{\mathcal{D}_r\}$
        \STATE \textbf{Input}: evaluator $\mathcal{E}$, $p_c$, $p_{\mathrm{LS}}$, repair budget $L$
        \STATE \textbf{Output}: offspring $h^{(1)},h^{(2)}$
        \IF{$\mathrm{rand}()<p_c$}
            \STATE Create $h^{(1)},h^{(2)}$ by two-point crossover on active genes
            \IF{$\mathrm{rand}()<p_{\mathrm{LS}}$}
                \FOR{each offspring $h$}
                    \STATE \textbf{Worst-gene repair:} Identify bottleneck $r^* = \arg\max_{r} (\hat{T}_{\mathrm{infer}} + \hat{T}_{\mathrm{trans}} + \hat{T}_{\mathrm{queue}})$
                    \STATE Evaluate up to $L/2$ alternative legal destinations for $r^*$
                    \STATE \textbf{2-opt swap:} Sample up to $L/2$ random request pairs $(r_a, r_b)$
                    \STATE Evaluate fitness $J$ after swapping destinations $h_{r_a} \leftrightarrow h_{r_b}$
                    \STATE Write the single best improving move (repair or swap) back into $h$
                \ENDFOR
            \ENDIF
        \ELSE
            \STATE Copy parents into $h^{(1)},h^{(2)}$
        \ENDIF
        \STATE \textbf{Return} $h^{(1)},h^{(2)}$
    \end{algorithmic}
\end{algorithm}

The local repair step targets batch-level interactions that are not visible from an
individual request alone. Several requests may each prefer the same fast endpoint, but
the combined assignment can increase its service duration and dominate the batch
makespan. Worst-gene repair moves work away from the request that currently limits
batch completion, while 2-opt swap can correct paired imbalance when two destinations
would be better exchanged. Mutation adds exploration by changing active genes with
probability $p_m$.

This repair policy is routing-specific: standard crossover mixes two parent
assignments through position-level inheritance, whereas CEMA-Router uses the
deployment objective to filter the local moves introduced after recombination. The
child passed to survivor selection already contains the accepted reassignment, so
useful local improvements can propagate without increasing the population size. This
bounded form matches online routing, where the router must improve placement quality
within a limited decision budget.

\subsection{Suitability for Batch-Level LLM Routing}

CEMA-Router fits batch-level LLM request routing for three reasons. First, its
chromosome directly matches the router-service output: one destination per request in
the current batch. Second, the evaluator uses the same deployment terms as the system
model, including profiled quality, explicit token cost, endpoint power, calibrated
batch throughput, network latency, and backlog. Third, the search is anytime, so the
router can return the best assignment found so far if the online decision budget is
exhausted.

The search cost is controlled by three explicit parameters: population size $P$,
generation count $G$, and the local-repair budget $L$. For each generation, the
dominant work is evaluating complete chromosomes under the four-term objective;
repair adds a bounded number of evaluator calls around sampled legal reassignments.
The router returns both the selected assignment and decision-time metrics, allowing
Section~\ref{sec:evaluation} to report routing overhead alongside objective value.

\section{Performance Evaluation}
\label{sec:evaluation}

\subsection{Experimental Setup}

\textbf{Datasets \& Workloads:} 
We utilize three datasets. The first is RouterBench~\cite{hu2024routerbench}, providing 36,497 trace-driven prompts evaluated across 11 LLMs with pre-computed ArmoRM~\cite{wang2024interpretable} quality scores and token costs. RouterBench is commonly used to evaluate contemporary multi-LLM routing approaches, including Eagle~\cite{zhao2024eagle}, MixLLM~\cite{wang2025mixllm}, and UniRoute~\cite{jitkrittum2025universal}. UniRoute is also our main quality--cost routing baseline, so RouterBench provides a shared basis for reproducible, isolated algorithmic comparison. For the real-world deployment, we evaluate 300 sampled questions from GSM8K~\cite{cobbe2021training} (scored by exact math match) and 600 from MMLU~\cite{hendrycks2021measuring} (scored by exact choice match) using a zero-shot chain-of-thought format.

\textbf{Testbed \& Infrastructure Profiles:}
For the controlled RouterBench stage, the 11 models are mapped to a simulated cloud--edge profile (Table~\ref{tab:routerbench-nodes}), with cloud nodes providing high throughput (up to 312 TFLOPS) but higher latency (40--60 ms) and token costs, whereas edge nodes offer lower power and latency (5--8 ms). 
For the testbed, we instantiate a heterogeneous six-node testbed (Table~\ref{tab:deployment-nodes}), hosting models ranging from SmolLM2-1.7B to Qwen2.5-14B. It is governed by real-time health checks, measured path TCP latencies (up to 197 ms for the remote V100), and calibrated saturated-batch throughput ($F_B$), dictating the continuous batching speedup.

\begin{table}[t]
    \caption{Cloud--edge profile used for RouterBench simulation}
    \label{tab:routerbench-nodes}
    \centering
    \scriptsize
    \resizebox{\columnwidth}{!}{%
    \begin{tabular}{@{}llllrp{34mm}@{}}
        \toprule
        Role & Class & TFLOPS & Power & Models \\
        \midrule
        Cloud & H100 & 312 & 300 W & GPT-4, Claude-v2\\
        Cloud & A100 & 312 & 300 W & GPT-3.5, Claude-v1, Claude-instant \\
        Cloud & A100 & 250 & 250 W & Llama-2-70B, Mixtral-8x7B \\
        Edge & A10 & 125 & 150 W & Yi-34B, Code-Llama-34B \\
        Edge & T4 & 65 & 70 W & WizardLM-13B, Mistral-7B \\
        \bottomrule
    \end{tabular}}
\end{table}

\begin{table}[t]
    \caption{Node configuration for real deployment.}
    \label{tab:deployment-nodes}
    \centering
    \scriptsize
    \begin{tabular}{@{}lllll@{}}
        \toprule
        Role  & GPU        & Model        & $F_1$  & $F_B$ \\
        \midrule
        Cloud & A100 40GB  & Qwen2.5-14B  & 4.02   & 287.5  \\
        Edge  & A40 24GB   & Phi-3.5-mini & 1.47   & 82.9   \\
        Edge  & A40 24GB   & Llama-3.1-8B & 1.70   & 167.0  \\
        Edge  & A30 24GB   & Qwen2.5-7B   & 2.23   & 158.6  \\
        Edge  & A30 24GB   & Mistral-7B   & 2.22   & 167.4  \\
        Cloud & V100 16GB  & SmolLM2-1.7B & 0.682  & 36.37  \\
        \bottomrule
    \end{tabular}
\end{table}

\subsection{Experimental Procedure}

We compare CEMA-Router against a recent LLM routing reference and several simple
deployment policies:
\begin{itemize}
    \item \textbf{CEMA-Router}: the proposed infrastructure-aware batch router,
    optimizing the normalized weighted objective over quality loss, monetary cost,
    energy consumption, and batch completion time.
    \item \textbf{UniRoute}~\cite{jitkrittum2025universal}: a quality-cost LLM router
    that clusters prompts by embedding similarity, estimates per-cluster error rates
    for each model, and scores each prompt--model choice using error rate plus a
    normalized cost term. In the RouterBench stage, we use $\lambda=0.05$ for its
    quality-cost score; in the deployment stage, UniRoute uses the same profiled
    quality source as CEMA-Router but does not score queueing, endpoint power, or
    batch makespan.
    \item \textbf{Round Robin}: a simple policy that assigns requests sequentially
    across currently available destinations, following a commonly evaluated routing
    baseline~\cite{yu2025efficient,jain2024intelligent}.
    \item \textbf{Least Loaded}: a simple policy that selects the destination with the
    lowest tracked outstanding work under the same Request Monitor state, analogous
    to shortest-queue load-aware routing~\cite{jain2024intelligent}.
    \item \textbf{Cloud Only} and \textbf{Edge Only}: placement policies that restrict
    routing to cloud or edge destinations, respectively, and then choose within that destination set; both are established cloud--edge comparison
    strategies~\cite{yu2025efficient,yang2024perllm}.
\end{itemize}

All methods are invoked through the same router service interface. Request Manager forms batches, the method switch selects the routing policy, and the selected assignment is evaluated using the same BatchEvaluator. In RouterBench stage, all methods use the trace-driven profile in Table~\ref{tab:routerbench-nodes}. In testbed stage, they use destination
profiles in Table~\ref{tab:deployment-nodes} and the quality context from Section~\ref{subsec:routing-context}. As noted in Sect.~\ref{subsec:routing-context}, the testbed study focuses on profiled-quality routing.

The real-deployment main comparison uses 10 runs per method on each workload.
Request-scalability experiments use batch sizes $R \in \{10,30,50\}$ with 10 runs per
method and dataset. To stress endpoint-pool scaling beyond the testbed
size, the controlled destination-scalability experiment fixes $R=30$ and varies the
available destinations as $M \in \{11,22,44,88\}$ over 30 independent RouterBench
batches by replicating the profiled node--model destination catalogue. The router
configuration uses a one-second batching interval, a maximum batch size of 128, and
default objective weights
$(\omega_Q,\omega_C,\omega_E,\omega_T)=(0.25,0.25,0.25,0.25)$. In the testbed,
the same weight vector is kept for implementation consistency across controlled and
deployed experiments, while the result discussion emphasizes quality, energy,
completion time, and decision overhead. The complete MA configuration and selection
rationale are reported in Sect.~\ref{subsec:hyperparameters}.

Performance is evaluated using the following metrics. Lower values are better for all reported metrics. \textbf{A) Weighted Objective}  the normalized weighted objective $J$ optimized for a given dataset, weight profile, and comparison setting. \textbf{B) Quality Loss}: raw profiled answer-quality loss, $Q_{\mathrm{raw}}$. \textbf{C) Monetary Cost}: explicit per-token charge, $C_{\mathrm{raw}}$. In the RouterBench simulations, monetary cost is computed using the \texttt{TOKEN\_COSTS} model-pricing table from the RouterBench benchmark code~\cite{hu2024routerbench}, which specifies per-token prompt and completion prices, and optional fixed per-request charges, in USD. The self-hosted deployment assigns zero external API price to all endpoints, so the cost term is retained for objective compatibility but is non-discriminative in the deployed experiments; the deployment result tables therefore focus on quality, energy, completion time, and routing overhead. \textbf{D) Energy Consumption}: estimated processing and communication energy, $E_{\mathrm{raw}}$, computed from the fixed node-power profile and modeled completion time. \textbf{E) Completion Time}: estimated batch completion time after dispatch, $T_{\mathrm{raw}}$. \textbf{F) Routing Overhead}: routing decision time measured separately from batch completion time.
The raw terms are reported alongside main objective $J$ so that the trade-off is auditable; a lower
$J$ can coexist with different movements in the raw metrics.

\subsection{Experimental Results}

\subsubsection{Hyperparameter Configuration}
\label{subsec:hyperparameters}

Table~\ref{tab:hyperparameters} summarizes preliminary tuning in heterogeneous
cloud--edge scenarios. 
\begin{table}[t]
    \caption{CEMA-Router hyperparameter configuration.}
    \label{tab:hyperparameters}
    \centering
    \scriptsize
    \renewcommand{\arraystretch}{1.08}
    \setlength{\tabcolsep}{2.5pt}
    \begin{tabular}{@{}>{\raggedright\arraybackslash}p{0.34\columnwidth}
                        >{\centering\arraybackslash}p{0.21\columnwidth}
                        >{\centering\arraybackslash}p{0.35\columnwidth}@{}}
        \toprule
        \textbf{Parameter} & \textbf{Tested range} & \textbf{Selected} \\
        \midrule
        Population size $P$ & $10$--$240$ & $30$ \\
        Generations $G$ & $20$--$120$ & $20$ \\
        Crossover rate $p_c$ & $0.8$--$0.9$ & $0.9$ \\
        Mutation rate $p_m$ & $0.05$--$0.2$ & $0.1$ \\
        Local-search mode & 3 variants & Probabilistic Lamarckian \\
        Local-search probability $p_{\mathrm{LS}}$ & $0$--$1$ & $0.5$ \\
        Objective weights $\boldsymbol{\omega}$ & - & $(0.25,0.25,0.25,0.25)$ \\
        \bottomrule
    \end{tabular}
\end{table}

Here, $R$ is the number of requests in a batch. Population sizes near $R$--$1.5R$
gave the best quality--runtime balance; larger populations improved quality by less
than 2\% while approximately doubling search time. With the default $R=30$, we chose
the lower end of this range, $P=30$, to limit online overhead. Most runs converged
within 5--23 generations, and generation 15 captured 93.5--98.8\% of the reduction
reached at generation 20, supporting $G=20$. This $P=30,G=20$ configuration achieved
mean fitness $J=0.079$ with 0.325 s decision time in the controlled RouterBench
evaluation. The
operator sweep selected stable $p_m=0.1$ and probabilistic Lamarckian search, which
gave the best mean fitness and reduced mean search time from 1.02 s to 0.18 s;
$p_c=0.9$, repair probability 0.5, tournament size 3, and an eight-trial repair budget
retain exploration within the online budget. No weight profile dominated every raw metric.

\subsubsection{Controlled RouterBench Results}
To evaluate CEMA-Router under repeatable trace-driven conditions, we first use RouterBench with the weighted objective. This stage fixes the prompt set,
RouterBench quality and cost data, and the five-node cloud--edge profile in
Table~\ref{tab:routerbench-nodes}. The controlled run uses $R=30$ and reports the mean over five sampled batches. The results are presented in
Table~\ref{tab:routerbench}. CEMA-Router ahieves the lowest weighted objective, while
also reducing cost, energy consumption, and completion time relative to UniRoute.

These results show that  jointly optimizing the
infrastructure-aware objective can produce routing outcomes beyond quality-cost routing.
The trace-driven evaluation isolates the scheduler, BatchEvaluator, and memetic search path, whereas 
the real deployment also evaluates real
endpoint health checks, path-latency measurement, request dispatch, and completion
monitoring. The trace-driven results further show that CEMA-Router incurs lower routing overhead, requiring 0.325 s on average compared with 0.481 s for the UniRoute baseline.

\begin{table}[t]
    \caption{Controlled RouterBench simulation. Weighted Objective, Normalized ($\widetilde{Q}$,$\widetilde{E}$, $\widetilde{C}$,$\widetilde{T}$), and raw results}
    \label{tab:routerbench}
    \centering
    \scriptsize
    \setlength{\tabcolsep}{2.5pt}
    \resizebox{\columnwidth}{!}{%
    \begin{tabular}{@{}lrrrrrrrrr@{}}
        \toprule
        Method & $J$ & $\widetilde{Q}$ & $Q_{\mathrm{raw}}$ &
        $\widetilde{C}$ & $C_{\mathrm{raw}}$ & $\widetilde{E}$ &
        $E_{\mathrm{raw}}$ & $\widetilde{T}$ & $T_{\mathrm{raw}}$ \\
        \midrule
        CEMA-Router & 0.087 & 0.186 & 0.370 & 0.097 & 0.0073 & 0.064 & 390.2 J & 0.000 & 0.456 s \\
        UniRoute & 0.212 & 0.022 & 0.317 & 0.061 & 0.0061 & 0.197 & 545.0 J & 0.569 & 1.441 s \\
        Edge Only & 0.369 & 1.000 & 0.627 & 0.000 & 0.0039 & 0.056 & 393.8 J & 0.419 & 1.188 s \\
        Least Loaded & 0.562 & 0.477 & 0.465 & 0.598 & 0.0260 & 0.646 & 1033.4 J & 0.526 & 1.347 s \\
        Round Robin & 0.562 & 0.477 & 0.465 & 0.598 & 0.0260 & 0.646 & 1033.4 J & 0.526 & 1.347 s \\
        Cloud Only & 0.817 & 0.268 & 0.400 & 1.000 & 0.0413 & 1.000 & 1438.4 J & 1.000 & 2.281 s \\
        \bottomrule
    \end{tabular}%
    }
\end{table}


\subsubsection{Real Deployment Results}
\begin{figure}[t]
    \centering
    \includegraphics[width=0.87\columnwidth]{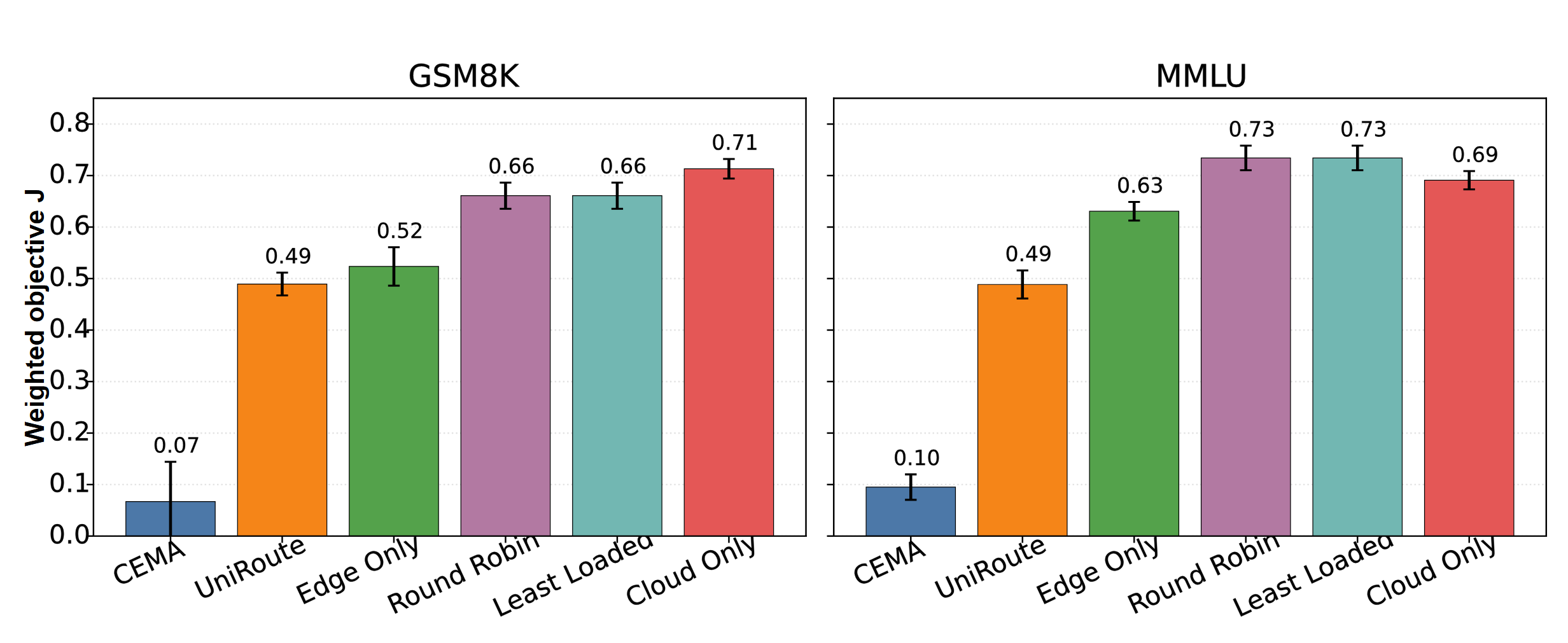}
    \caption{Real-deployment main comparison on GSM8K and MMLU under the weighted
    infrastructure-aware objective (R=30, 10 runs).}
    \label{fig:main-comparison}
\end{figure}

Figure~\ref{fig:main-comparison} and Table~\ref{tab:real-main} exhibit the real-deployment comparison on GSM8K and MMLU, exercising the full runtime paths defined in Sect.~\ref{sec:system}. On both datasets, CEMA-Router achieves substantially lower values in the weighted objective $J$.

\begin{table}[t]
    \caption{Real-deployment CEMA vs. UniRoute main results.}
    \label{tab:real-main}
    \centering
    \scriptsize
    \begin{tabular}{@{}llrrrrr@{}}
        \toprule
        Data & Method & $J$ & $Q_{\mathrm{raw}}$ & $E_{\mathrm{raw}}$ & $T_{\mathrm{raw}}$ & Dec. \\
        \midrule
        GSM8K & CEMA & 0.0669 & 0.3670 & 12928 J & 2.443 s & 1.059 s \\
        GSM8K & UniRoute & 0.4894 & 0.3362 & 19541 J & 3.351 s & 0.523 s \\
        MMLU & CEMA & 0.0951 & 0.3643 & 12947 J & 2.342 s & 1.757 s \\
        MMLU & UniRoute & 0.4886 & 0.2746 & 16087 J & 3.351 s & 1.612 s \\
        \bottomrule
    \end{tabular}
\end{table}

These real-deployment values show where CEMA-Router's additional routing context
matters. On GSM8K, the weighted objective decreases by 86.3\%, while energy
consumption decreases by 33.8\% and completion time decreases by 27.1\%. On MMLU,
the same pattern holds: the objective decreases by 80.5\%, energy decreases by
19.5\%, and completion time decreases by 30.1\%. Both routers use the same
profiled-quality input, but CEMA-Router also accounts for endpoint load, network
delay, saturated batch throughput, and energy profiles. A semantically strong
destination can have an unfavorable queuing or energy profile for the current
batch; batch-level assignment allows CEMA-Router to choose a lower-straggler or
lower-energy placement because of the weighted objective.

\subsubsection{Metric Trade-off Analysis}
\begin{figure}[t]
    \centering
    \includegraphics[width=0.88\columnwidth]{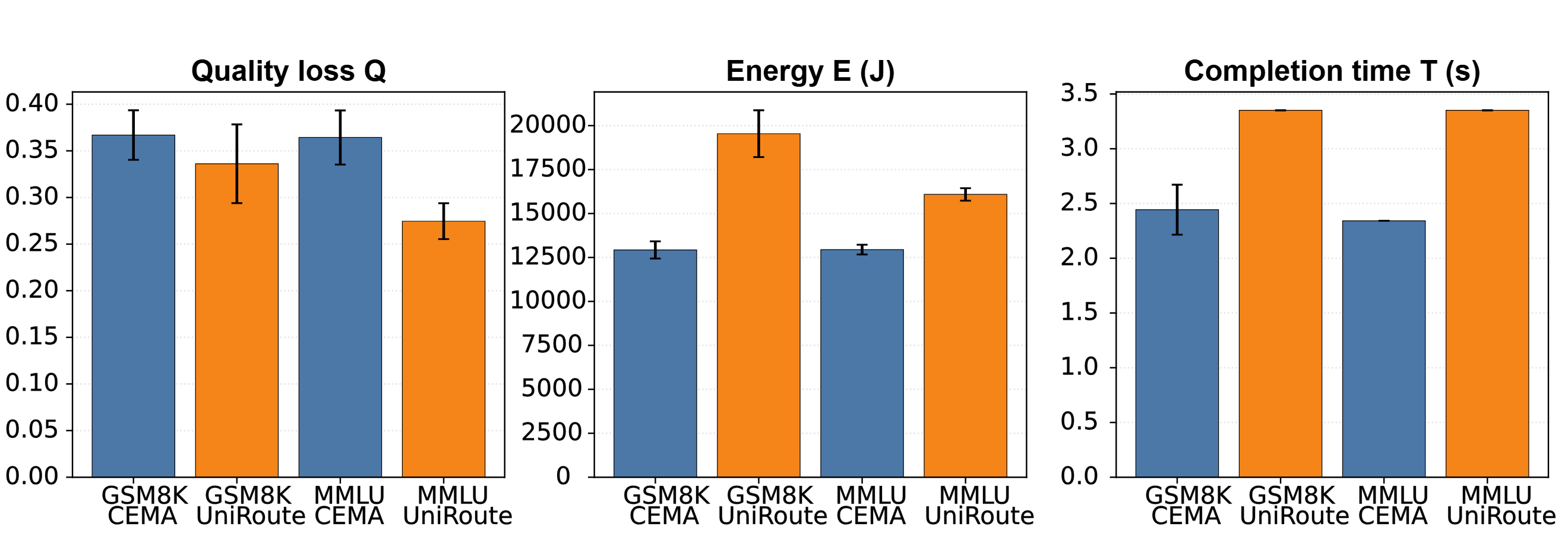}
    \caption{Raw metric trade-offs between CEMA-Router and UniRoute in the real
    deployment. It shows the different raw-metric
    priorities induced by model-level and infrastructure-aware routing.}
    \label{fig:tradeoff}
\end{figure}

Figure~\ref{fig:tradeoff} decomposes the CEMA--UniRoute trade-off. CEMA-Router routes
more work toward endpoints that reduce energy consumption and completion time under the
deployment profile. UniRoute instead follows its model-level quality-cost objective. The result is an
objective-level improvement, with different raw metrics moving in different
directions.

This decomposition clarifies the role of each comparator. UniRoute is a strong
reference for model-level quality-cost routing. CEMA-Router adds queuing, calibrated
batch throughput, endpoint power profiles, and straggler completion time to the
decision, exposing how deployment terms change the selected assignment under the
configured objective.

\subsubsection{Request Scalability Results}

\begin{figure}[t]
    \centering
    \includegraphics[width=0.99\columnwidth]{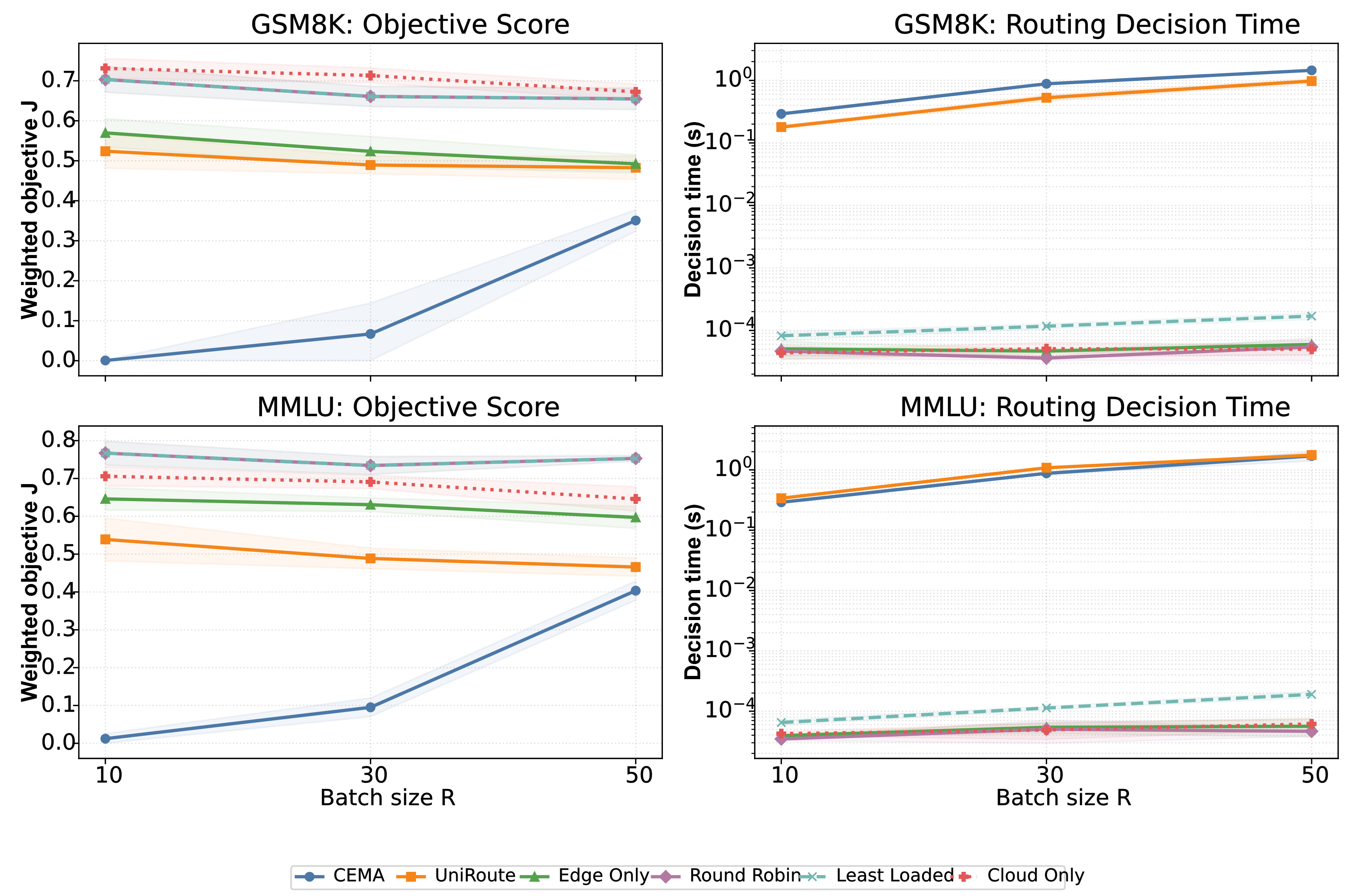}
    \caption{Request scalability in the real deployment (10 runs).}
    \label{fig:scalability}
\end{figure}

Figure~\ref{fig:scalability} and Table~\ref{tab:scalability} show request scalability
for $R=10, 30, 50$. This experiment stresses the connection between
Section~\ref{subsec:request-management}, which admits larger batches, and
Section~\ref{sec:design}, which searches over longer assignment
chromosomes. CEMA-Router preserves a lower weighted objective (lower is better) than
UniRoute across both datasets and all batch sizes. The same experiment shows
that its decision time grows with batch size and is higher than UniRoute in the real
deployment.

\begin{table}[t]
    \caption{Request scalability summary for CEMA and UniRoute.}
    \label{tab:scalability}
    \centering
    \scriptsize
    \begin{tabular}{@{}llrrrr@{}}
        \toprule
        Dataset & $R$ & CEMA $J$ & UniRoute $J$ & CEMA dec. & UniRoute dec. \\
        \midrule
        GSM8K & 10 & 0.0005 & 0.5239 & 0.437 s & 0.179 s \\
        GSM8K & 30 & 0.0669 & 0.4894 & 1.324 s & 0.529 s \\
        GSM8K & 50 & 0.3506 & 0.4826 & 2.175 s & 0.978 s \\
        MMLU & 10 & 0.0123 & 0.5389 & 0.435 s & 0.339 s \\
        MMLU & 30 & 0.0951 & 0.4886 & 1.313 s & 1.083 s \\
        MMLU & 50 & 0.4035 & 0.4658 & 2.548 s & 1.758 s \\
        \bottomrule
    \end{tabular}
\end{table}

\subsubsection{Destination Scalability Results}

\begin{figure}[t]
    \centering
    \includegraphics[width=0.92\columnwidth]{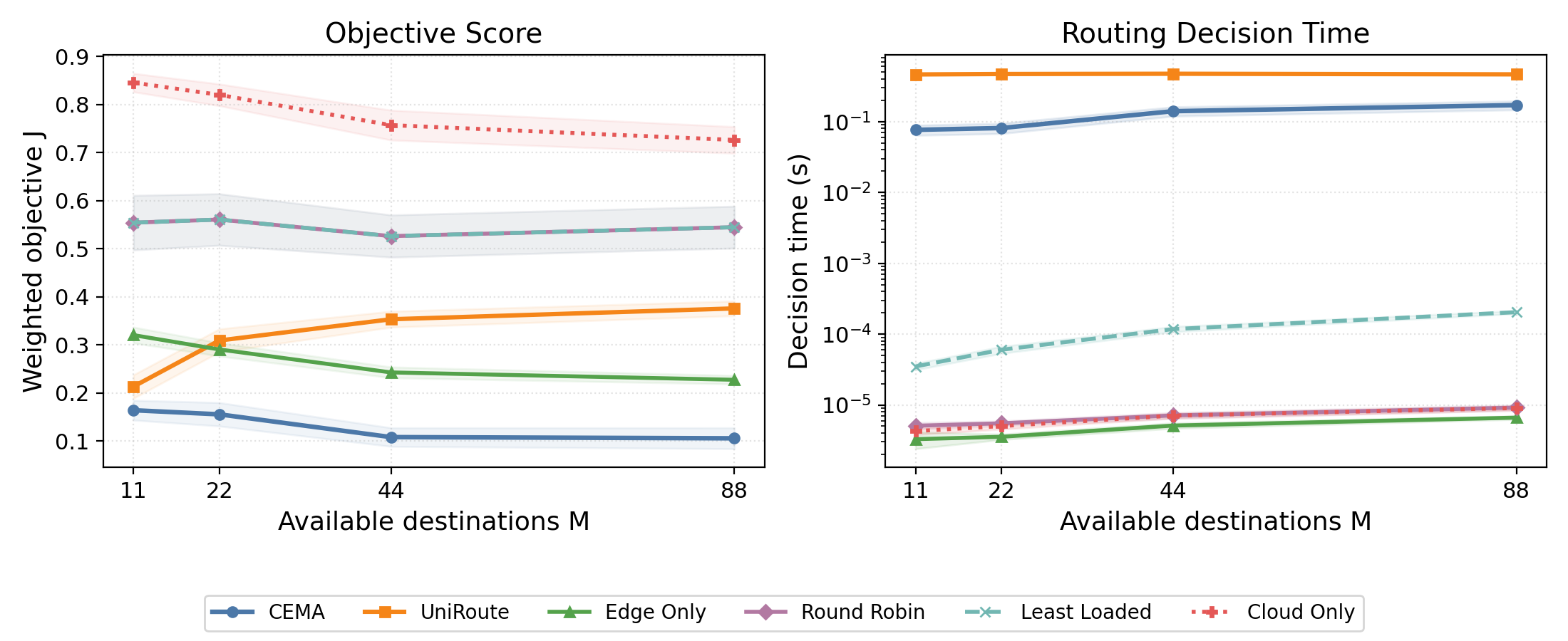}
    \caption{Controlled RouterBench destination scalability with increasing available
    destinations. Shaded bands denote 95\%
    confidence intervals over 30 independent batches.}
    \label{fig:destination-scalability}
\end{figure}

Figure~\ref{fig:destination-scalability} shows controlled RouterBench destination scalability for $M=11, 22, 44, 88$ available destinations with $R=30$. 
CEMA-Router achieves the lowest weighted objective
at all scales, improving from $J=0.164$ at $M=11$ to
$J=0.106$ at $M=88$, whereas UniRoute increases from $J=0.214$ to $0.376$. 
Its
bounded search budget keeps decision time sub-second, increasing from
0.077 s to 0.171 s as $M$ grows, compared with approximately 0.46--0.47~s for UniRoute.

\subsection{Discussion}
CEMA-Router demonstrates the value of treating LLM routing as a joint infrastructure-aware assignment problem rather than considering model quality or system efficiency in isolation. Across the controlled RouterBench study and the six-endpoint deployment, CEMA-Router achieves lower weighted objective values than UniRoute and the deployment-oriented heuristics. These results indicate that combining request-specific quality estimates with node throughput, network conditions, energy profiles, and dynamic backlog information can identify operating points that are not visible to quality--cost or load-only policies. Nevertheless, the proposed weighted objective represents a configurable deployment preference rather than universal dominance across every raw metric.
\par
The raw results expose the underlying optimisation trade-off. CEMA-Router reduces estimated energy consumption and straggler-dominated completion time by directing requests away from congested or energy-intensive destinations, but it may accept a higher quality loss than the quality-oriented UniRoute policy. In the controlled evaluation, it can also incur a slightly higher token cost. Thus, the preferred assignment depends on application requirements: latency- or energy-sensitive pervasive services may tolerate a small quality reduction, whereas safety-critical or accuracy-sensitive applications may require a minimum-quality constraint. The objective weights allow operators to express these preferences, while a quality threshold can be incorporated when degradation beyond a specified level is unacceptable. Because all endpoints in the self-hosted deployment have zero external API cost, the deployed comparison primarily reflects the trade-off among quality, energy, and completion time.
\par
The results also illustrate the benefit of batch-level placement. A per-query policy may independently assign several requests to the same apparently attractive endpoint without considering the contention created by their combined execution. CEMA-Router instead evaluates complete assignments using inherited backlog, predicted request length, network delay, and calibrated batch throughput. Its load-aware seed initially distributes longer requests according to expected finishing time, while the repair and swap operators subsequently correct bottlenecks at the batch level. This advantage is most pronounced when endpoint capabilities and request sizes are heterogeneous and several requests compete for the same high-performance destination. When batches are small or one endpoint dominates all metrics, the additional benefit of joint placement is expected to decrease.
\par
This additional coordination introduces routing overhead. The controlled simulation primarily operates on precomputed quality and infrastructure profiles, whereas the deployed pipeline additionally performs online prompt embedding, FAISS-based quality retrieval, profiler access, context construction, and router-service communication. These operations, together with the memetic search, explain why deployment decision time is higher and more workload-dependent than controlled simulation time (Fig.~\ref{fig:scalability} and ~\ref{fig:destination-scalability}). For small batches, the overhead remains within the batching interval; for larger batches, it can exceed that interval and offset part of the execution-time saving. CEMA-Router is therefore most beneficial when the avoided queuing and inference delay exceeds its routing cost. Practical optimisations include caching prompt embeddings and retrieved profiles, vectorising fitness evaluation, applying adaptive population and generation budgets, terminating search when improvement stalls, and overlapping routing for the next window with current-batch execution.

\section{Conclusion and Future Work}

This paper presented CEMA-Router, an infrastructure-aware batch router for LLM inference in cloud--edge systems. The system seamlessly integrates request batching, dynamic backlog monitoring, and RAG-profiled quality signals with a customized memetic assignment optimizer. Across controlled RouterBench traces and a real six-node GSM8K/MMLU deployment, CEMA-Router outperforms the quality-cost-centric UniRoute and standard deployment heuristics. Notably, it reduces the overall infrastructure-aware objective by over 80\%, effectively cutting energy consumption by up to 33.8\% and batch completion time by 27.1\%. The decision-time measurements identify practical operating regimes for population-based batch routing and support scalable deployment mechanisms such as early-stop budgets, asynchronous planning, and adaptive batching. Overall, the results demonstrate that treating infrastructure heterogeneity, response quality, energy consumption, and continuous-batching queueing as joint runtime signals is crucial. By moving beyond isolated model selection, CEMA-Router provides a vital step toward sustainable and latency-bounded pervasive AI.

Future work will extend CEMA-Router toward more dynamic pervasive environments, including mobility-driven wireless variation, endpoint churn, and online profile refresh. We also plan to integrate continuous bandwidth and power telemetry from deployed edge devices to evaluate routing stability under broader long-term workloads.

\bibliographystyle{IEEEtran}
\bibliography{references}

\end{document}